\documentclass[aps,prc,superscriptaddress,twoside,twocolumn,nofootinbib,%
              showpacs,floatfix]{revtex4-1}

\usepackage{amsmath,amssymb}
\usepackage{graphicx,bm}
\usepackage{slashed}

\begin{document}

\title{\boldmath Model-independent determination of the parity of $\Xi$ hyperons}

\author{K. Nakayama}%
\email{nakayama@uga.edu}
\affiliation{Department of Physics and Astronomy, University of Georgia,
Athens, GA 30602, USA}
\affiliation{Institut f\"ur Kernphysik and Center for Hadron Physics, Forschungszentrum
J\"ulich, D-52425 J\"ulich, Germany}

\author{Yongseok Oh}%
\email{yohphy@knu.ac.kr}
\affiliation{Department of Physics,
Kyungpook National University, Daegu 702-701, Korea}
\affiliation{Asia Pacific Center for Theoretical Physics, Pohang, Gyeongbuk 790-784, Korea}

\author{H. Haberzettl}%
\email{helmut@gwu.edu}
\affiliation{Institute for Nuclear Studies and Department of Physics,
The George Washington University, Washington, DC 20052, USA}

\date{\today}

\begin{abstract}
Based on reflection symmetry in the reaction plane, it is shown that measuring
the transverse spin-transfer coefficient $K_{yy}$ in the $\bar{K} N \to K\Xi$
reaction directly determines the parity of the produced cascade hyperon in a
model-independent way as $\pi_\Xi^{} =K_{yy}$, where $\pi_\Xi^{} =\pm 1$ is the parity.
This result based on Bohr's theorem provides a completely general, universal
relationship that applies to the entire hyperon spectrum. A similar expression
is obtained for the photoreaction $\gamma N \to K K \Xi$ by measuring both the
double-polarization observable $K_{yy}$ and the photon-beam asymmetry $\Sigma$.
Regarding the feasibility of such experiments, it is pointed out that
the self-analyzing property of the $\Xi$'s can be invoked,
thus requiring only a polarized nucleon target.
\end{abstract}


\maketitle


The spectrum of multi-strangeness hyperons is largely unknown and much is yet
to be explored. For example, the flavor SU(3) symmetry leads to the expectation
that the number of $\Xi$ resonances is equal to the number of non-strange
baryon resonances, i.e., the nucleon and $\Delta$ resonances. However, the
compilation of particle data found in the Particle Data Group Review
(PDG)~\cite{PDG10} shows that to date only a dozen $\Xi$'s have been discovered
compared to about 40 non-strange baryon resonances. Furthermore, only two of
them, $\Xi(1318)$ and $\Xi(1530)$, have four-star status according to the PDG.
One of the reasons for this situation is that the $\Xi$ hyperons, being
particles with strangeness $S=-2$, can only be produced via indirect processes
from the nucleon, with very small production yields that makes them difficult
to measure. Moreover, the situation is exacerbated by the lack of facilities
that can produce anti-kaon beams. As a result, nothing of significance
regarding $\Xi$ resonances has been added to the PDG listings during the last
two decades~\cite{PDG10}. The situation is going to change very soon with the
availability of the anti-kaon beam at the Japan Proton Accelerator Research
Complex (J-PARC) facility which has started its operation just recently. Since
the anti-kaon has strangeness $S=-1$, hyperons with $S=-2$ ($\Xi$) can be
produced directly in reactions such as $\bar{K} N \to K \Xi$ with sufficiently
large yields. Indeed, the study of multi-strangeness hyperons is one of the
major parts of the physics programs at J-PARC~\cite{Nagae07,Nagae08}.
Furthermore, the $\overline{\rm P}$ANDA Collaboration has also proposed an
investigation of the $\bar{p} p \to \bar{\Xi} \Xi$ reaction at the Facility for
Antiproton and Ion Research (FAIR)~\cite{SP04,PANDA09}. Also, the CLAS
Collaboration at the Thomas Jefferson National Accelerator Facility (JLab) has
established the feasibility of investigating $\Xi$ spectroscopy via
photoproduction reactions like $\gamma p \to K^+ K^+ \Xi^-$ and $\gamma p \to
K^+ K^+ \pi^-\Xi^0$~\cite{CLAS04d,PDGN05,CLAS06d}. The first dedicated
experiment for these reactions has been carried out and the data for the total
and differential cross sections as well as the $K^+K^+$ and $K^+\Xi^-$
invariant mass distributions for the $\gamma p \to K^+ K^+ \Xi^-$ reaction have
been reported in Ref.~\cite{CLAS07b}. To our knowledge, this is the first data
set measured for the exclusive production of the $\Xi$ in photon-nucleon
scattering.

Theoretical studies of the production of $\Xi$ hyperons also started only recently.
For example, the production mechanisms for $\Xi$ photoproduction,
i.e., $\gamma p \to K^+ K^+ \Xi^-$ was investigated in Refs.~\cite{NOH06,MON11}
and recent works for the $\bar{K}N \to K\Xi$ reaction were reported in
Refs.~\cite{SKL11,SST11}.

The investigation of multi-strangeness baryons is expected to shed light on our
understanding of the structure of baryons and it will allow us to distinguish
various phenomenological models of the baryon-mass spectrum
\cite{Oh07,CIK81,CI86,PR07,VXN11}.  Knowing the parity quantum number, in
particular, is of crucial importance in baryon spectroscopy since it heavily
depends on the internal structure of the baryon. However, the experimental
extraction of the spin-parity quantum numbers is very difficult. The parity
quantum number of the $\Xi$ ground state, $\Xi(1318)$, in particular, has not
been measured yet but is assigned to be positive in the PDG compilation based
on the quark-model predictions~\cite{PDG10}. Therefore, given this uncertain
situation, reliable experimental determinations of the quantum numbers of the
$\Xi$ ground state and its resonances are important and timely and of
particular interest for the experimental programs at facilities that can
produce cascades, like J-PARC and others.

There are some earlier efforts to determine the spin-parity quantum numbers of
a cascade resonance, in particular, of $\Xi^*(1820)$, through an analysis of the
moments of its decay products~\cite{Minn77,TDDG78,BBBB87}. The procedure of
Ref.~\cite{BBBB87} permits the determination of both spin and parity. However,
it is limited to resonances above threshold with odd relative orbital angular
momentum between the decay products.

In this article, we show an alternative, completely model-independent and
universal way of determining the parity of any $\Xi$ hyperon with an
arbitrary spin. This is based on Bohr's theorem~\cite{Bohr59} which is a
consequence of the invariance of the transition amplitude under rotation and
parity inversion and, in particular, reflection symmetry in the reaction plane.
To this end, we consider the reaction $\bar{K} N \to K \Xi$, where the $\Xi$
hyperon has spin $j$. This is is one of the reactions that will be studied at
J-PARC. For completeness, we also consider the parity determination of $\Xi$
via the photoproduction reaction $\gamma N \to K K \Xi$, which can be performed
at JLab.

While completely general, the foremost practical use of the method lies with
the ground state of the hyperon, as we shall argue below, in the paragraph
following Eq.~(\ref{eq:7}). Therefore, we first consider the case of a spin
$j=1/2$ cascade hyperon and show explicitly that the transverse spin-transfer
coefficient in $\Xi$ production in the $\bar{K} N$ scattering is directly
related  to the parity of the $\Xi$ hyperon. We will then generalize the
results to the case of a $\Xi$ hyperon with an arbitrary spin $j$. The most
general spin-structure of the reaction amplitude, consistent with symmetry
principles, for the process $\bar{K}(q) + N(p) \to K(q') + \Xi(p')$, where the
arguments $q$, $p$, $q'$, and $p'$ stand for the four-momenta of the respective
particles, is given by
\begin{subequations}\label{eq:1}
\begin{align}
\hat{M}^+ & = M_0 + M_2 \, \bm{\sigma}\cdot\hat{\bm{n}}_2^{}~,\\
\hat{M}^- &  = M_1 \, \bm{\sigma}\cdot\hat{\bm{n}}_1^{} + M_3 \, \bm{\sigma}\cdot\hat{\bm{q}}~,
\end{align}
\end{subequations}%
for positive and negative parity $\Xi$ ($\hat{M}^+$ and $\hat{M}^-$),
respectively. Here, $\bm{\sigma}=(\sigma_1,\sigma_2,\sigma_3)$ is the Cartesian
vector made up of the three Pauli matrices $\sigma_i$, with indices $1,2,3$
corresponding to spatial axes $x,y,z$. The unit  vectors $\hat{\bm{n}}_1^{}$
and $\hat{\bm{n}}_2^{}$ are defined as $\hat{\bm{n}}_1^{} \equiv
(\bm{q}\times\bm{q}')\times\bm{q} / |(\bm{q}\times\bm{q}')\times\bm{q} |$ and
$\hat{\bm{n}}_2^{} \equiv \bm{q}\times\bm{q}' / |\bm{q}\times\bm{q}' |$,
respectively.

Without loss of generality, we may choose the coordinate systems such that
$\bm{q}$ is along the positive $z$-axis and $\hat{\bm{n}}_2^{}$ along the
positive $y$-axis. Then $\hat{\bm{n}}_1^{}$ is the unit vector along the
positive $x$-axis. The plane containing the vectors $\bm{q}$ and
$\hat{\bm{n}}_1^{}$ is the reaction plane and $\hat{\bm{n}}_2^{}$  is
perpendicular to that plane.

 The reaction amplitudes in Eq.~(\ref{eq:1}) can be summarily written as
\begin{equation}
\hat{M} = \sum_{m=0}^3 M_m^{} \sigma_m^{} \ ,
\label{eq:2}
\end{equation}
where in addition to the three Pauli matrices  $\sigma_i^{}$ ($i=1,2,3$),
$\sigma_0^{}$ here is the 2$\times$2 unit matrix. For a positive-parity $\Xi$,
$M_1=M_3=0$ and $\hat{M}$ reduces to $\hat{M}^+$, and for a negative-parity
$\Xi$, $\hat{M}$ reduces to $\hat{M}^-$ because $M_0=M_2=0$. Expressing
amplitudes utilizing $\hat{M}$ of Eq.~(\ref{eq:2}), the unpolarized cross
section is given by
\begin{equation}
\frac{d\sigma}{d\Omega} \equiv \frac{1}{2} \,\mbox{Tr}\left( \hat{M}\hat{M}^\dagger \right)
= \sum_{m=0}^3 |M_m|^2
\label{eq:3}
\end{equation}
and the (diagonal) spin-transfer coefficient $K_{ii}$ ($i=1,2,3$) is obtained as
\begin{align}
\frac{d\sigma}{d\Omega}K_{ii} & \equiv \frac{1}{2} \,\mbox{Tr}\left(
\hat{M}\sigma_i^{} \hat{M}^\dagger\sigma_i^{} \right) \nonumber \\
& = |M_0|^2 + |M_i|^2 - \sum_{k\ne i} |M_k|^2 \ .
\label{eq:4}
\end{align}
In terms of the cross sections, the spin-transfer coefficient $K_{ii}$ is given
by
\begin{equation}
K_{ii} =
\frac{\left[d\sigma_i^{}(++) + d\sigma_i^{}(--)\right]- \left[d\sigma_i^{}(+-)+d\sigma_i^{}(-+)\right]}
{\left[d\sigma_i^{}(++) + d\sigma_i^{}(--)\right] + \left[d\sigma_i^{}(+-)+d\sigma_i^{}(-+)\right]} \ ,
\label{eq:4a}
\end{equation}
where $d\sigma_i^{}$ stands for the differential cross section with the
polarization of the target nucleon and of the produced cascade along the
$i$-direction. The first and second $\pm$ arguments of $d\sigma_i^{}$ indicate
the parallel ($+$) or anti-parallel ($-$) spin-alignment along the
$i$-direction of the target nucleon and produced cascade, respectively.

In general, $K_{ii}$ depends on the energy and scattering angle.
However, from Eqs.~(\ref{eq:3}) and (\ref{eq:4}), it follows immediately that
$K_{yy}$ is constant and that it provides the parity $\pi_{\Xi}^{}$ of $\Xi$, \textit{viz.}
\begin{equation}
\pi_{\Xi}^{} = \pm 1=K_{yy}~ ,
\label{eq:5}
\end{equation}
where the sign directly corresponds to positive or negative parity.
This result is a direct consequence of the spin structures of the reaction amplitudes for
positive and negative parity $\Xi$ as exhibited in Eq.~(\ref{eq:1}), which, in turn,
is a consequence of reflection symmetry in the reaction plane.

The above results can be straightforwardly generalized to a $\Xi$ with an
arbitrary spin $j$ by invoking Bohr's theorem~\cite{Bohr59} written in the
form~\cite{Satchler}
\begin{equation}
\pi_{fi}^{} = (-1)^{M_f - M_i}~.
\label{eq:6}
\end{equation}
Here, $\pi_{fi}^{}$ denotes the product of the intrinsic parities of all the
particles in the initial ($i$) and final ($f$) states, while $M_{(i/f)}$ stands
for the sum of the spin projection quantum numbers of the initial/final state
particles along the axis perpendicular to the reaction plane, i.e.,
$\hat{\bm{n}}_2^{}$ or the $\hat{y}$-axis. For the reaction in question,
$\pi_{fi}^{} = \pi_{\Xi}^{}$, and thus
\begin{equation}
\pi_{\Xi}^{}=  (-1)^{M_f - M_i} =K_{yy}  ~ .
\label{eq:7}
\end{equation}
The results given in Eqs.~(\ref{eq:5}) and (\ref{eq:7}), therefore, directly
determine the parity of the produced $\Xi$ hyperon.

To discuss the feasibility of this determination, the present results show that
to obtain the parity of $\Xi$, one needs to measure the double-polarization
observable $K_{yy}$ which is usually extremely challenging experimentally. The
task simplifies considerably if the weak decay modes of the hyperon can be
separated from the strong ones, which is particularly true for the ground state
where strong decays are absent altogether. One may then employ the fact that
the cascade states are self-analyzing under weak decays~\cite{textbooks}, thus
requiring only a polarized nucleon target to determine the polarization of the
$\Xi$'s. Such a polarized target may be available at J-PARC in the foreseeable
future~\cite{Sawada}, which would make this experiment possible for the
spin-1/2 ground state of $\Xi$. The feasibility of such an experiment hinges on
the cross-section yield with a polarized nucleon target, which should be
smaller than the unpolarized cross section by roughly a factor of 10 if one
assumes a typical degree of polarization of $\sim 20\%$ of the target nucleon.
Since the unpolarized cross section for $K^- p \to K^+ \Xi^-$ is of the order
of $10\,\mu$b/sr around $\sqrt{s} \sim 2
$~GeV~\cite{BGLOR66,LRSY66,TS67,TFHL68,BMPT68,DBHM69}, one might expect
cross-section yields of the order of $1\,\mu$b/sr with the polarized nucleon
target.

It should be mentioned that, in principle, for $j=1/2$ the parity of
the cascade resonance may also be determined by measuring
single-polarization observables, namely, the target-nucleon asymmetry, $T_i$,
and the recoil-cascade polarization, $P_i$,
\begin{subequations}\label{eq:10}
\begin{align}
\frac{d\sigma}{d\Omega}T_i &\equiv \frac{1}{2} \,\mbox{Tr}\left( M\sigma_i M^\dagger \right)
                                              = 2\,\mbox{Re}\left[M_0 M^*_i \right]
                                             + 2\,\mbox{Im} \left[M_j M^*_k \right]\,,  \\
\frac{d\sigma}{d\Omega}P_i &\equiv \frac{1}{2}\,\mbox{Tr}\left( M M^\dagger \sigma_i \right)
                                               = 2\,\mbox{Re}\left[M_0M^*_i \right]
                                               - 2\,\mbox{Im}\left[M_j M^*_k \right]\,,
\end{align}
\end{subequations}
where the subscripts $(i,j,k)$ runs cyclically, i.e., (1,2,3), (3,1,2), (2,3,1).
Then, with the amplitudes given by Eq.~(\ref{eq:1}), it follows immediately that
\begin{subequations}
\begin{align}
\frac{d\sigma}{d\Omega}\left( T_y + P_y\right) & = 4\,\textrm{Re}\left[ M_0M^*_2 \right]~,
\label{eq:10a}\\
\frac{d\sigma}{d\Omega}\left( T_y  - P_y\right) & = 0 \ ,
\label{eq:10b}
\end{align}
\end{subequations}
for positive-parity cascade and
\begin{subequations}
\begin{align}
\frac{d\sigma}{d\Omega}\left( T_y  + P_y\right) & = 0 \ ,
\label{eq:11a} \\
\frac{d\sigma}{d\Omega}\left( T_y - P_y\right) & = 4\,\textrm{Im}\left[ M_3M^*_1 \right] \ ,
\label{eq:11b}
\end{align}
\end{subequations}
for negative-parity cascade.
The equations here reveal that if the measured combination $T_y+P_y$ is different
from zero the parity of the cascade is positive; conversely, if the combination
$T_y - P_y$ is different from zero, the parity is negative.
Here, it should be noted that the usefulness of these expressions hinges on how
reliably the respective right-hand sides of Eqs.~(\ref{eq:10a}) and (\ref{eq:11b})
can be determined to be different from zero experimentally, which may not be
possible if any one of the amplitudes $M_i\ (i=0,1,2,3)$ is too small.
This magnitude problem aside, this experiment would be easier to set up because
it requires only measuring the single-polarization observables, $T_y$ and $P_y$.
The determination of the double-polarization observable, $K_{yy}$, by contrast,
while requiring a more complex experimental setup, is free of any potential
magnitude problem.

The parity of the cascade hyperon may also be determined in a model-independent
way in the photoproduction reaction $\gamma N \to K K \Xi$ that will be studied
at JLab~\cite{Guo}. Since kaons are spin-zero particles, in this case one can
simply make use of the results derived in Ref.~\cite{NL04}.%
\footnote{Details of the derivation of the spin-structure of
 the photoproduction amplitudes used in Ref.~\cite{NL04}, especially those
 involving a negative-parity baryon, can be found in Ref.~\cite{NL05}.}
One finds, in particular, among the various spin observables and combinations
of spin observables for this reaction that can be used in principle to determine the
parity of $\Xi$, the transverse spin-transfer coefficient with the unpolarized photon
beam $K_{yy}$ and the photon-beam asymmetry $\Sigma$ are related to the parity
of the $\Xi$ resonance by
\begin{equation}
 \pi_{\Xi}^{} =\frac{K_{yy}}{\Sigma}~.
\label{eq:9}
\end{equation}
Obviously, here the measurements of spin observables, especially the
double-polarization observable $K_{yy}$, are more challenging than in hadronic
reactions due to much smaller cross-section yields.

The relations found here for the $\Xi$ hyperon can also be applied to the
parity determination of the $\Omega$ hyperon in the reactions $\bar{K}N \to
KK\Omega$ and $\gamma N \to KKK \Omega$. Because the production yields are much
smaller for the $\Omega$ hyperons than those for the $\Xi$ hyperons, the
required measurements of polarization observables in $\Omega$ production would
be much more difficult. At any rate, our results given in Eqs.~(\ref{eq:5}),
(\ref{eq:7}), and (\ref{eq:9}) can be used to determine the parity of $\Omega$
hyperons in principle. However, because of the presence of an additional kaon in $\Omega$
production, $\pi_\Xi^{}$ in these relations should be replaced by
$-\pi_\Omega^{}$.

In summary,  we have shown that, based on reflection symmetry in the reaction
plane, the parity of a $\Xi$ hyperon with an arbitrary spin can be directly
determined in a model-independent, universal manner by measuring the transverse
spin-transfer coefficient $K_{yy}$ in the $\bar{K}N \to K\Xi$ reaction that
will be studied at the J-PARC facility. Our result is particularly relevant for
the ground state of the $\Xi$ since in this case one may exploit the fact that
the $\Xi$ is self-analyzing under weak decays. In principle, however, our
theoretical result applies to the entire cascade spectrum. The parity of the
cascade hyperon may also be determined in the photoproduction reaction $\gamma
N \to KK\Xi$, provided one can measure the transverse spin-transfer coefficient
with the unpolarized photon beam and the beam asymmetry with linearly polarized
photons. We also mention that since the respective quantities $K_{yy}$ and
$K_{yy}/\Sigma$ for both types of experiments need to be equal to known
constants (i.e., $\pi_\Xi^{}=\pm1$), apart from providing the parity of $\Xi$,
measurements of these quantities also provide some lower limits for the
systematic errors of such experiments. Regarding the practical feasibility of
such experiments, we mention that the self-analyzing feature of the $\Xi$'s
will help if their weak decay modes can be measured to determine the
polarization of these $\Xi$ hyperons. Finally, we add that the present
discussions can also be applied to the parity determination of $\Omega$
hyperons.

\acknowledgments

We are grateful to Lei Guo, Johann Haidenbauer, Wooyoung Kim, 
Wim Kloet, Shin'ya Sawada, and Igor Strakovsky
for fruitful discussions. This work was partly supported by the National
Research Foundation of Korea funded by the Korean Government (Grant No.\
NRF-2011-220-C00011). The work of K.N. was also supported partly by the FFE
Grant No. 41788390 (COSY-058).

\end{document}